\documentclass[multphys,vecphys]{svmult}

\usepackage{makeidx}         % allows index generation
\usepackage{graphicx}        % standard LaTeX graphics tool
\usepackage{multicol}        % used for the two-column index
\usepackage[bottom]{footmisc}% places footnotes at page bottom
\makeindex             % used for the subject index
\begin{document}

\title*{Eta Car through the eyes of interferometers}

\titlerunning{Eta Car through the eyes of interferometers}
\authorrunning{Chesneau, O. et al.} 

\author{O.~Chesneau\inst{1}, R.~van Boekel\inst{2}, T.~Herbst\inst{2}, P.~Kervella\inst{3}, M.~Min\inst{4},  L.B.F.M.~Waters\inst{4}, Ch.~Leinert\inst{2}, R. Petrov\inst{5} and G. Weigelt\inst{6}}
\institute{$^1$ Observatoire de la C\^{o}te d'Azur-CNRS-UMR 6203, Dept. Gemini,
Avenue Copernic, 06130 Grasse, France
\texttt{Olivier.Chesneau@obs-azur.fr}
\\$^2$ Max-Planck-Institut für Astronomie, K\"onigstuhl 17, 69117 Heidelberg, Germany
\\$^3$ LESIA, CNRS-UMR 8109, Observatoire de Paris-Meudon, 5 place Jules Janssen, 92195 Meudon Cedex, France
\\$^4$ Sterrenkundig Instituut `Anton
Pannekoek', Kruislaan 403, 1098 SJ Amsterdam, The Netherlands
\\$^5$ Universit{\'e} de Nice-Sophia Antipolis, Parc Valrose, 06108 Nice, France
\\$^6$ Max-Planck-Institut für Radioastronomie Auf dem Hügel 69 53121 Bonn, Germany
}
\maketitle

\begin{abstract}
The core of the nebula surrounding Eta Carinae has recently been observed with VLT/NACO, VLTI/VINCI, VLTI/MIDI and VLTI/AMBER in order to 
spatially {\it and} spectrally constrain the warm dusty environment and the
central object. Narrow-band images at 3.74~$\mu$m and 4.05~$\mu$m
reveal the structured butterfly-shaped dusty environment close to
the central star with an unprecedented spatial resolution of about
60~mas. VINCI has resolved the present-day stellar wind of Eta
Carinae on a scale of several stellar radii owing to the spatial
resolution of the order of 5~mas ($\sim$ 11~AU). The VINCI observations show that the object
is elongated with a de-projected axis ratio of approximately 1.5.
Moreover the major axis is aligned with that of the large bipolar
nebula that was ejected in the 19th century. Fringes have also
been obtained in the Mid-IR with MIDI using baselines of 75m. A
peak of correlated flux of 100~Jy is detected 0.3"
south-east from the photocenter of the nebula at 8.7~$\mu$m is
detected. This correlated flux is partly attributed to the central
object but it is worth noting that at these wavelengths, virtually all the 0.5" x 0.5" central area can generate detectable fringes
witnessing the large clumping of the dusty ejecta. These
observations provide an upper limit for the SED of the central
source from 3.8~$\mu$m to 13.5~$\mu$m and constrain some
parameters of the stellar wind which can be compared to Hillier's model. Lastly, we present the great potential of the AMBER instrument to study
the numerous near-IR emissive lines from the star and its close vicinity. In particular, we discuss its ability 
to detect and follow the faint companion.
\end{abstract}

\section{Introduction}
\label{sec:1}

Eta Carinae is one of the best studied but least understood
massive stars in our galaxy (\cite{davidson2}). Eta Car
is classified as a Luminous Blue Variable (LBV); a short lasting phase of hot star
evolution characterized by strong
stellar winds and instabilities leading to possible giant eruptions. 
Eta Car offers a unique opportunity to observe the 
consequences of such a giant events with the two historical eruptions in the 1840s and 1890s.

The large bipolar nebula surrounding the central object, known as the
``Homunculus'', was formed during the first one, while some fainter and less extended structures were attributed to the lesser 1890 event\footnote{although by that time the dust absorption was such that the opacity around the central object was such that the amplitude of this event is not really constrained}. Despite numerous observations and theoretical studies, the cause of the two outbursts remains
unknown.
 Currently, the Homunculus lobes span a
bit less than 20$"$ on the sky (or 45000 AU at the system distance
of 2.3 kpc) and are largely responsible for the huge infrared
luminosity of the system. The central source is therefore deeply embedded and suffers from high extinction due to the nebular dust.

Improved spatial resolution observations have often been  the key for the 
progress in our understanding of this emblematic embedded object.
The central source has been studied by speckle interferometry
techniques in visible light, which revealed a complex knotty structure (\cite{weigelt86}, \cite{falcke}). Originally, three remarkably
compact objects between 0.1$"$ and 0.3$"$ northwest of the
star were isolated (the so-called BCD {\it Weigelt blobs}, the
blob A being the star itself). Other similar but fainter objects
have since been detected (\cite{weigelt95}; \cite{davidson1}). They are found to be surprisingly bright
ejecta moving at low speeds ($\sim$50 km.s$^{-1}$). They
belong to the equatorial regions close to the star; their
separation from the star is typically 800 AU.

Eta Car has been systematically observed with Hubble Space Telescope (HST) instruments (\cite{davidson1}, GHRS; \cite{morse}, \cite{king}, WFPC2/NICMOS; \cite{smith2004a}, \cite{smith2004b}, ACS/HRC among many others) and in particular the Space Telescope Imaging Spectrograph (STIS) since the beginning of 1998 (\cite{ishibashi}, \cite{smith2003b}, \cite{verner}, \cite{gull}). With 0.1 arcsec angular resolution and a spectral resolving power of 5000, the central point-like source can be studied more or less independently from the extended nebulosity and the nebular structures can be dissected (cf. contribution from T. Gull in these proceedings). The central source could be separated from the Weigelt blobs allowing a careful study of the central object by Hillier et al. (\cite{hillier}). 
The mostly reflective nebula also allowed an indirect study
by STIS of the stellar wind from several points of view at different latitudes
in the nebulae by means of reflected P Cygni absorption in Balmer lines (\cite{smith2003b}). The authors convincingly
prove the asphericity of the wind, suggesting an enhanced polar wind mass-loss.

Eta Car was observed with the Infrared Space
Observatory (ISO, \cite{morris}). The ISO
spectra indicated that a much larger amount of matter should be
present around Eta Car in the form of cold dust than previously
estimated. Observations with higher spatial resolution by Smith et
al. (\cite{smith2002a}, \cite{smith2003a}) showed a complex but organized
dusty structure within the three inner arcseconds. They showed
that the dust content in the vicinity of the star is relatively limited and
claimed that the two polar lobes should contain the large mass of
relatively cool dust necessary to explain the ISO observations (but see \cite{dekoter}).
The mechanism
required to form the two gigantic lobes of gas and dust remains poorly understood.

One of the main limitations of HST observations is probably that its instruments are
mostly restricted to the optical domain. The Homunculus is a dusty nebula dominated by reflected starlight at optical wavelengths and even in K band, the scattered light from the central object is far from being negligible. This makes it problematic to really look through all this material without being affected by the diffuse light. Lower extinction in the IR allows us to look inside the Homunculus to study embedded structures, provided that the spatial resolution is sufficient to study them.

In the following, we present the recent results obtained by the impressive gain in spatial resolution provided by the VLT with the NACO instrument (Sect.1) and the VLTI with the VINCI (Sect.2) and MIDI (Sect.3) instruments. In Sec.4 we present the potential of AMBER observations for detecting and observing the companion of Eta Car.

% Always give a unique label
% and use \ref{<label>} for cross-references
% and \cite{<label>} for bibliographic references
% use \sectionmark{}
% to alter or adjust the section heading in the running head

\section{NACO observations: the inner dusty nebula}
\label{sec:2}

The interferometric observations presented in the following sections were complemented with broad- and
narrow-band observations taken with the NAOS/CONICA (NACO) imager
installed on VLT UT4 (Kueyen), equipped with an adaptive optics (AO)
system. These observations are described extensively in van Boekel et al. (\cite{vanboekel}) and Chesneau et al. (\cite{chesneau}).

The diffraction limit (defined by the Point Spread Function, PSF of the telescope) of a 8 meter telescope at 3.8~$\mu$m
is about 100~mas. At this wavelength, the NACO adaptive optics sufficiently
corrects the atmospheric seeing, routinely providing a
Strehl ratio approaching 0.5. A careful deconvolution procedure can
improve the spatial resolution to about 50-80 mas, i.e. close to the diffraction limit of K band images usually obtained with lower strehl performances and more affected by scattered light (see below).
These NACO images represent the highest-resolution images
of Eta Car in the K and L bands presently available (shown in the Figure 1). They resolve much of
the sub-arcsecond structures.

The NACO observations offer the
opportunity to bridge the gap between existing observations and the interferometric data
obtained with very high resolution but sparse UV coverage. They are also a great help for interferometric observations employing single mode fibers for which the field of view (FOV) is strongly limited.

\begin{figure}
\centering
\includegraphics[width=12cm]{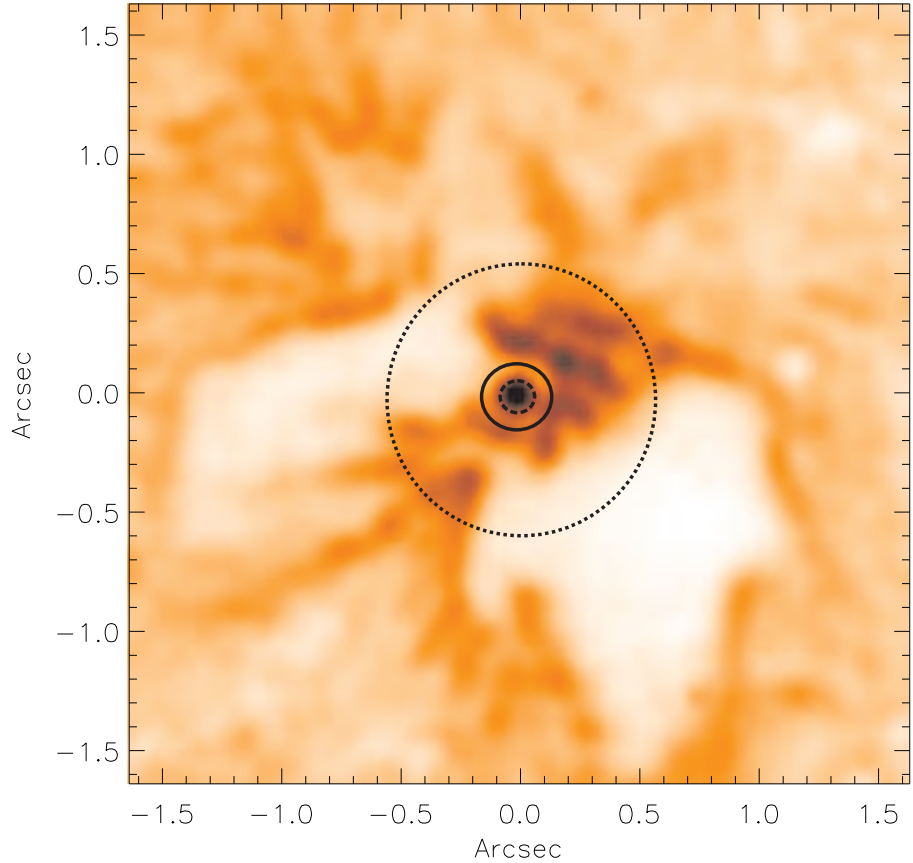}
\caption{NACO Pf$\gamma$ deconvolved image. The
  resolution achieved is of the order of 60~mas. To reduce the contrast, the image I$^{1/4}$ is shown.
  The FOV of the different interferometric instruments of the VLTI with UTs and ATs are shown: 2.2$\mu$m with ATs and 10$\mu$m with UTs in solid line, 2.2$\mu$m with UTs in dashed line, 10$\mu$m with ATs in dotted line.}
\label{fig:1}       % Give a unique label
\end{figure}

Dust plays a key role in the study of Eta Car. It intervenes in
every observation as strong and patchy extinction is
frequently invoked as an important process in explaining the
photometric variability of Eta Car. However, the exact nature
and location of dust formation/destruction sites has never been observed.
In the close vicinity of the central object, dust is still present in large quantities and even the K band images
are contaminated by the intense scattered light which decreases strongly only in L band.

A 3.8~$\mu$m narrow-band deconvolved image is shown is Fig. 1. This image illustrates the difficulty of observing such a complex and extended object with long baseline
interferometers. The key difficulty is that all spatial
structures at the scale of one Airy disk contribute to the
interferometric signal, especially the single mode AMBER and VINCI interferograms. 
AMBER and VINCI are sensitive to the 1-20 mas structures at the center of the FOV and are "contaminated" by everything inside a complex 100-150~mas patch with the UTs and inside a 400-600~mas patch with ATs at 2.2~$\mu$m. This means that the visibility and phase can be contaminated by the closest regions of the "Weigelt complex". If we want to be fully able to interpret AMBER data, we would have to accurately control the pointing and do mosaicing to explore the inner 30-150~mas region. The AMBER capability to disentangle between narrow (from blobs) and broad (from central star) emission lines owing to its high spectral resolution (R=10000) should be used to provide the necessary astrometric information for assessing the pointing quality (see Sect.~\ref{sec:5}).

For MIDI observations, the situation is even worse since in this case the dust emission flux is several times larger than that of the central star and virtually all the nebula is bright. The airy patterns of UTs and ATs delimit a region of 0.25 and 1.25 arcsec respectively. One avantage though, is the availability of a 3" FOV to simultaneously get fringes from extended regions;
i.e., not to be restricted to the FOV of a single Airy disk. 

\section{VINCI observations: the evidence for rotation}
\label{sec:3}

Rotation is an intrinsic property of all
stars, which definitely cannot be neglected in the case of early spectral types.
The most obvious consequence is the geometrical
deformation that results in a
radius larger at the equator than at the
poles. Another well established effect,
known as gravity darkening or the von
Zeipel effect, is that both the surface
gravity and emitted flux decrease from
the poles to the equator. Although well
studied in the literature, such effects of
rotation have rarely been directly tested
against observations.

It must be pointed out that in the case of Eta Car, the wind density is such that the true photosphere is not visible. Therefore, the consequences of rotation
can only be indirectly observed through their effects on the dense wind of this star.
Until the indirect observations of Smith et al. (\cite{smith2003b}), the common thought was that centrifugal forces favor mass-loss in the equatorial plane. We will see in the following that interferometry is the most appropriate tool to detect directly the asymmetry if the central source.

The VINCI observations of Eta Car are described in van Boekel et al. (\cite{vanboekel}).
The two 35 cm siderostats and the instrument VINCI were used to obtain interferometric measurements at baselines ranging from 8 to 62 m in length. The observations were carried out in the first half of 2002 in four different nights, and again in early 2003. The baselines used, have a ground length of 8, 16, 24, and 66 m respectively. In particular observations with the 24 m baseline cover a wide range of projected baseline orientations.

VINCI observations provide information on the K continuum from the central object, the flux from emission lines being limited to less than a few percent of the total flux in this band. As mentioned in the previous section, the extended FOV of siderostats includes, in addition to the central source (representing 57\% of the observed flux), the regions from the 'Weigelt complex' within the dotted curve in Fig.1. AMBER should be used to evaluate potential measurement biases.

\begin{figure}
\centering
\includegraphics[width=10cm]{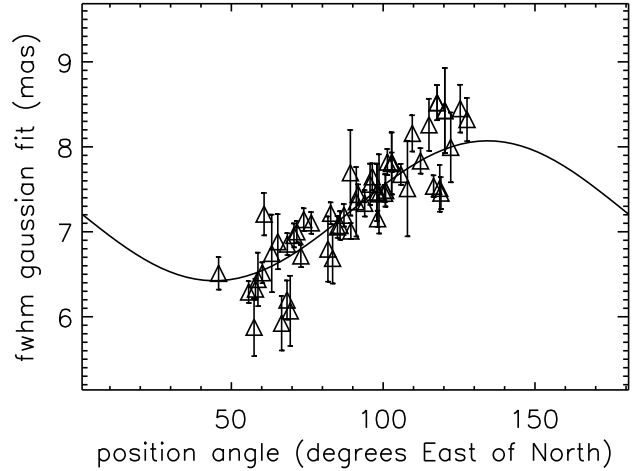}
\caption{Variation of FWHM fitted to the visibilities measured with VINCI as a function of projected orientation of the 24 m baseline. The solid line gives the best fit to the measurements, assuming a 2D Gaussian shape of the source at each projected baseline orientation. The amplitude of the size variations gives a ratio of major to minor axis of 1.25$\pm$0.05. The major axis has a position angle of 134$^\circ$$\pm$7$^\circ$ East of North.}
\label{fig:1}       % Give a unique label
\end{figure}

VLTI/VINCI observations clearly
resolve this central object; its size can now
be measured to be 5 mas at 2~$\mu$m corresponding
to 10 AU at the distance of Eta Car.
This is much larger than the stellar
photosphere so that we must be observing
an optically thick wind. The radiation is dominated
by free-free emission and electron
scattering; the emerging spatial intensity profile is
determined by the mass-loss rate and the
wind clumping factor. The intensity profile measure
with the VLTI breaks the degeneracy
between these two parameters in
previous modeling efforts; mass loss rate
and clumping factor can be derived separately
from the combination of HST/STIS spectroscopy and the interferometric
data.
These
observations are consistent with the presence of a star which
has an ionized, moderately clumpy stellar wind with a mass loss
rate of about 1.6x10$^{-3}$ M$_{\odot}$ yr$^{-1}$. This star-plus-wind
spherical model, developed by Hillier et al. (\cite{hillier}), is also
consistent with the HST STIS observations of the central object.

A second important conclusion from
the VLTI data is that the central object is
not spherically symmetric, the star is elongated with a
de-projected axis ratio of about 1.5. Moreover, its major axis is
aligned with that of the Homunculus. These VLTI observations provide a direct measurement of the wind geometry proposed by Smith et al. (\cite{smith2003b}).
The alignment on all scales means that the
1840 outburst looks like a scaled-up version
of the present-day wind, and that this
wind is stronger along the poles than in
the equatorial plane.  As Dwarkadas \& Owocki (\cite{dwarkadas}) showed, the radiation pressure
in these massive stars is stronger in
the polar regions because of the von
Zeipel effect. The wind is primarily controlled by this phenomenon and not by the local gravity.

\section{MIDI observations: dusty clumps, everywhere!}
\label{sec:4}

The MIDI recombiner attached to the VLTI is the only instrument
that is able to provide  sufficient spatial and spectral
resolution in the mid-infrared to disentangle the central
components in the Eta Car system from the dusty environment.
The Hillier model suggests a flux level of 200-300~Jy at 10~$\mu$m and 10-15 mas diameter of the star plus wind at this wavelength.

Eta Car was observed with MIDI with the UT1 and the UT3 telescopes
during commissioning and guaranteed time observations. These observations are reported in Chesneau et al. (\cite{chesneau}).
Virtually all the capabilities of MIDI were used during these 4h observations:
single-dish imaging during acquisition, spatially resolved long slit spectroscopy, undispersed
mapping of the correlation pattern and, finally, dispersed fringes. All data were taken within the small, but indeed of great interest, 3" FOV of the instrument.

\begin{figure}
\centering
  \includegraphics[width=5.8cm]{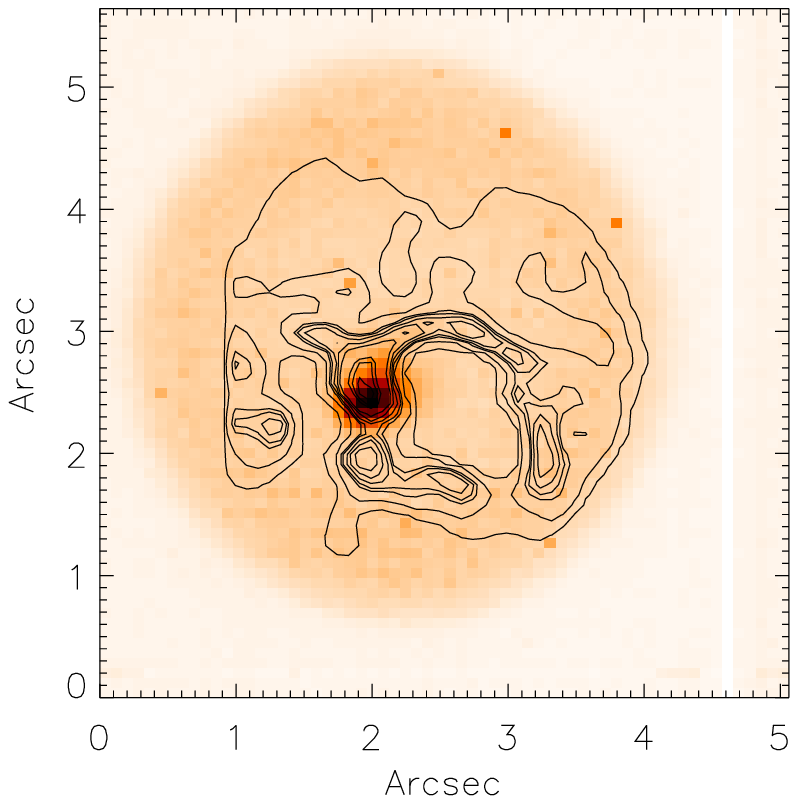}
  \includegraphics[width=5.8cm]{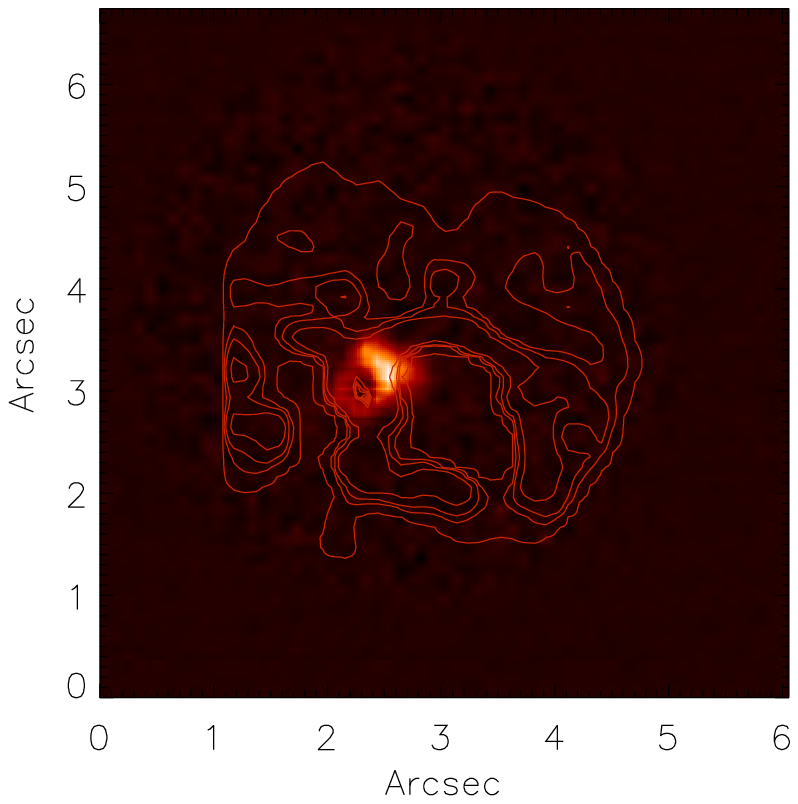}
\caption{Left, the figure shows the signal fluctuations within the
MIDI acquisition FOV. The external regions are dominated by the detector noise
and the internal regions by the tunnel and sky background
fluctuations. The signal from the fringes is strong and centered
on the position of the star as seen in the Fig.1 image. The contour plot represents the contours of
the deconvolved MIDI 8.7 $\mu$m acquisition image. Right, the
noise pattern and the fringe pattern from a calibrator have been
subtracted from the previous figure in order to show the extended
fringe signal. The vertical orientation is approximately parallel to the bipolar nebula (PA=138$^\circ$).}
\label{fig:2}       % Give a unique label
\end{figure}

The spatial distribution of the fringes
detected by MIDI with the 8.7~$\mu$m filter is shown in
Fig.\ref{fig:2}. The peak of the
fringes is localized at the position of the star
itself but an extended halo is also visible
in the Weigelt complex about 0.4-0.6" northwest from
the star. This is the confirmation that highly compressed material
emitting strongly at 8.7~$\mu$m exists in this region.
The fringes at the location of the Weigelt
blobs are definitely more extended than a single PSF FWHM at
8.7~$\mu$m (220~mas). 

This implies that
in the equatorial Weigelt region a fraction of the dust is
embedded in clumps with a typical size smaller than 10-20~mas
(25-50~AU) within a total extent of about 1000 AU. Nevertheless, this
correlated flux represents only a few percent of the total flux at
these locations. It must be pointed out that only a few scans with
fringes have been recorded during this commissioning measurement
and the lowest detectable fringe signal visible in
Fig.~\ref{fig:2} is about 20~Jy. With more integrated frames it should be possible to see that all the Weigelt complex indeed
can generate fringes. 

\begin{figure}
\centering
  \includegraphics[width=12.cm]{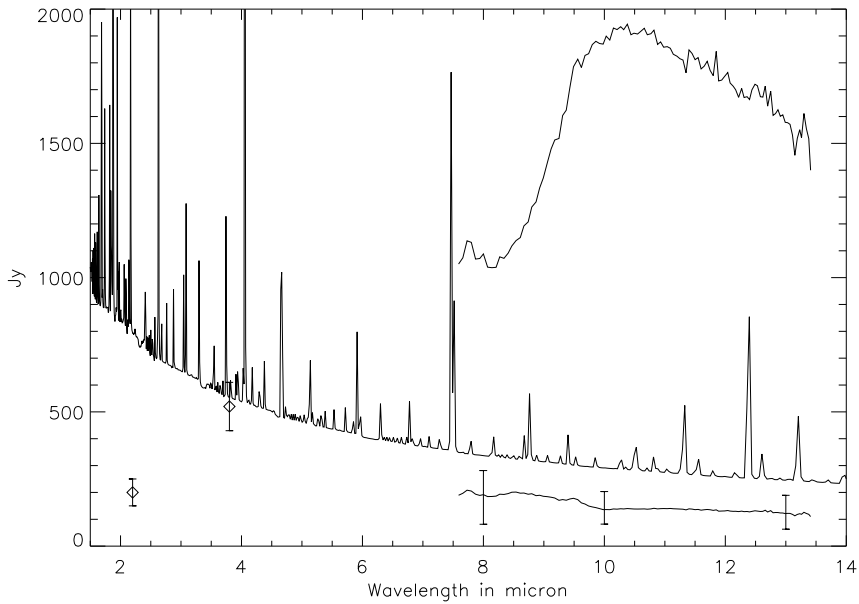}
\caption{Spectral energy distribution from Hillier's model compared with the photometry obtained from an airy disk centered on the star (upper curve) and the extracted correlated flux corrected from the expected visibities of the model (lower cuver). The near-IR photometry in the K and L band are also indicated. }
\label{fig:3}       % Give a unique label
\end{figure}

Dispersed fringes were also obtained which reveal a correlated flux of about
100~Jy situated 0.3" south-east of the photocenter of the
nebula at 8.7~$\mu$m, which corresponds with the location of the
star as seen in NACO images. This correlated flux is partly
attributed to the central object, and these observations together with the VINCI ones provide
an upper limit for the SED of the central source from 2.2~$\mu$m
to 13.5~$\mu$m (Fig.~\ref{fig:3}). 

The 74m baselines were roughly perpendicular to the main axis of the
nebula and the putative rotation axis of the prolate star itself. This means that the baselines were oriented perpendicular
to the main stellar axis, in which direction the star is smaller, corresponding
to a maximum correlated flux. Hence, the MIDI measurements can be
considered as an upper limit of the correlated flux observable
from the star. Despite large error bars, the MIDI correlated flux is obviously below the model predictions. We also note that the MIDI data were acquired at the periastron passage of the faint companion (still undetected), and no emission lines in the extracted spectra are visible compared to the model\footnote{The whole Eta Car spectrum is dominated by strong emission lines which disappear during a short time at periastron passage. Despite the low spectral resolution of MIDI (R=30 with the prism), we can clearly say that these lines were absent for these observations. The emission lines are for instance well detectable in HD316285, a twin of Eta Car from the spectral point of view, with the same spectral resolution.}.

Nine spectrally dispersed observations from the nebula itself at PA=318 degree, i.e. in the
direction of the bipolar nebula within the
MIDI field of view of 3" were also extracted. A large amount of corundum
(Al$_2$O$_3$) is discovered, peaking at 0.6-1.2"
south-east from the star, whereas the dust content of the Weigelt
blobs is dominated by silicates. These observations are extensively discussed in Chesneau et al. (\cite{chesneau}).

The garanteed time observations presented here were intended to judge the feasibility of MIDI observations of Eta Car. The observations, conducted with UTs only, demonstrated the interest of such a scientific program but also the difficulty to extract the central star signal from the dust one. The results were obtained with only one baseline and at periastron passage of the companion (see following section). In particular, a possible explanation of the low correlated flux observed, compared with the model of Hillier,
could be the presence of the extended wind-collision zone between the
central star and the fainter companion, commonly thought to be the place of an intense dust formation. The garanteed time with UTs has been completed and the star is now being offered for open time observations. It would be of great interest, now that the orbital phase is near apastron, to undertake a more ambitious program with UTs taking advantage of the fact that AMBER is ready to perform contemporary observations. Some test observations are also planned to check the feasibility of a large observing program with ATs. Their large FOV could prevent the extraction of pertinent information for such a complex object but they also offer a wealth of baselines that could provide a good uv coverage.

\pagebreak

\section{AMBER observations: revealing the binary?}
\label{sec:5}

The original detection of a 5.52 year period in Eta Carinae
in the spectroscopic and near-infrared photometric data of
Damineli (\cite{damineli96}) has been confirmed by later observations
(\cite{damineli00}, \cite{davidson3}, \cite{whitelock}). The
existence, mass, and orbit of a companion and its possible impact on the
behavior of the primary are still strongly disputed (e.g. \cite{davidson3}; \cite{corcoran}, \cite{feast}, \cite{pittard}; \cite{duncan}).

The deduced parameters of the orbit from X-ray observations by Corcoran
et al. (\cite{corcoran}) imply that the periastron and apastron distances are
roughly 1.5 AU and almost 30 AU, respectively (see also \cite{pittard}). At closest approach, the secondary is well
embedded, deep within the dense wind of the B-star primary. 

The first observations of Eta Car with AMBER appear promising (see Petrov et al., this volume). Already, absolute and differential visibilities, differential and closure phases were obtained at medium (R=1500) and high (R=10000) spectral resolution. Even with this impressive amount of information, the spatial interpretation of this complex (and time variable) object remains limited by the poor uv coverage. We advocate a spectroscopic approach based on the study of the numerous near-infrared lines formed at different spatial locations, viewing the central star through different angles and experiencing different excitation mechanisms (cf. in particular Smith et al. 2002b, Fig.14 and Fig.15).

Near-infrared emission lines are unique diagnostics of the geometry and kinematics the wind of Eta Car and for studying in its close vicinity. The infrared spectrum is a strong function of the position in Eta Car's nebula, with molecular hydrogen  and  [FeII] tracing more easily cold or collisionally excited material formed in the circumstellar outside the inner 1" region (\cite{smith2002b}, \cite{smith2005}). The HeI$\lambda$10830 line, variable with the orbital cycle ((\cite{damineli96}) and emitted from some equatorial ejecta (\cite{smith2002b}) is of particular interest. A. Damineli has been monitoring HeI$\lambda$10830, Pa$\gamma$ and Pa$\delta$ over the past several years and showed that this line is very sensitive to the orbital motion of the companion (\cite{damineli96}, \cite{damineli97}, see also the web site of A. Damineli http://www.etacarinae.iag.usp.br/). 

Smith (\cite{smith2002b}) was able to marginally resolve the emission from the Weigelt blobs, allowing their combined spectrum to be separated from the central star at IR wavelengths for the first time. During the first AMBER observations, it was observed that
the narrow emission lines are offset from the star's position by 0.2 to 0.4 arcsec, while the broad lines from the star wind appear when the telescope is accurately pointed. Thus, it should be possible to separate these blobs from the star by extracting segments along the slit on either side of the star's position. To the NW the emission is dominated by the blobs, and to the SE the star dominates. 

The seach for any indices that are able to constrain the
companion is now being intensively conducted.
By studying the high and low state of excitation from the emission lines of Weigelt blobs,
Verner et al. (\cite{verner}) found consistent results  with an O supergiant or a Wolf-
Rayet (W-R) star. Falceta-Goncalves et al. (\cite{falceta}) recently studied the accumulation of material in the wind-wind 
collision zone in an attempt to conciliate X-rays and optical observations. They also considered the formation rate of dust in the collision zone and came to interesting conclusions, complementary to those in the MIDI and NACO observations by Chesneau et al. (\cite{chesneau}). Their Fig.3, showing the expected tail of gas and dust created in the collision zone, illustrates the potential complexity of the object, which will be soon intensively observed with AMBER.

AMBER is particularly well suited to study the wind-wind collision zone and should detect it more easily than the companion itself: the emitting zone is large (i.e. well resolved), the wind-wind collision zone might be bright near periastron, and the radial velocities encountered for this kind of phenomenon can probe further the potentially complex geometry. The simultaneous use of three telescopes is indeed the best advantage of these AMBER observations. The observations of the famous WR-O binary system $\gamma^2$~Velorum (cf. Petrov et al., this volume) show the potential of AMBER for such a study and the preliminary results are very promising. 

%\pagebreak

%%%%%%%%%%%%%%%%%%%%%%%% referenc.tex %%%%%%%%%%%%%%%%%%%%%%%%%%%%%%
% sample references
% "physics"
%
% Use this file as a template for your own input.
%
%%%%%%%%%%%%%%%%%%%%%%%% Springer-Verlag %%%%%%%%%%%%%%%%%%%%%%%%%%

%
% BibTeX users please use
% \bibliographystyle{}
% \bibliography{}

%%%%%%%%%%%%%%%%%%%%%%%%%%%%%%%%%%%%%%%%%%%%%%%%%%%%%%%%%%%%%%%%%%%%%%  }

%%%%%%%%%%%%%%%%%%%%%%%%%%%%%%%%%%%%%%%%%%%%%%%%%%%%%%%%%%%%%%%%%%%%%%

\printindex
\end{document}